\documentclass[traditabstract]{aahyf} 
\usepackage{graphicx}

\usepackage{indentfirst}
\newcommand{\ud}{\mathrm{d}}

\begin{document}
   \title{On the afterglow from the receding jet of gamma-ray burst}

   \author{Xin Wang
      \inst{  }
   \and Y. F. Huang
      \inst{  }
   \thanks{\email{hyf@nju.edu.cn}}
   \and S. W. Kong
      \inst{  }
      }
   \institute{Department of Astronomy, Nanjing University, Nanjing 210093,
   China}

   \date{Received month day 2009 / Accepted month day 2009}

  \abstract{
According to popular progenitor models of gamma-ray bursts, twin
jets should be launched by the central engine, with a forward jet
moving toward the observer and a receding jet (or the counter jet)
moving backwardly. However, in calculating the afterglows, usually
only the emission from the forward jet is considered. Here we
present a detailed numerical study on the afterglow from the
receding jet. Our calculation is based on a generic dynamical
description, and includes some delicate ingredients such as the
effect of the equal arrival time surface. It is found that the
emission from the receding jet is generally rather weak. In radio
bands, it usually peaks at a time of $t \geq 1000$ d, with the peak
flux nearly 4 orders of magnitude lower than the peak flux of the
forward jet. Also, it usually manifests as a short plateau in the
total afterglow light curve, but not as an obvious rebrightening as
once expected. In optical bands, the contribution from the receding
jet is even weaker, with the peak flux being $\sim 23$ magnitudes
lower than the peak flux of the forward jet. We thus argue that the
emission from the receding jet is very difficult to detect. However,
in some special cases, i.e., when the circum-burst medium density is
very high, or if the parameters of the receding jet is quite
different from those of the forward jet, the emission from the
receding jet can be significantly enhanced and may still emerge as a
marked rebrightening. We suggest that the search for receding jet
emission should mostly concentrate on nearby gamma-ray bursts, and
the observation campaign should last for at least several hundred
days for each event. }


\titlerunning{Emission from the receding jet of GRB}
\authorrunning{Xin Wang et al.}

\maketitle
%

\section{Introduction}

Thanks to the discovery of X-ray, optical and radio afterglows of
gamma-ray bursts (GRBs), it is now clear that most GRBs are situated
at cosmological distances (Costa et al. 1997; van Paradijs et al.
1997; Frail et al. 1997). A lot of progresses have been achieved
during the past decade (Piran 2004; M\'esz\'aros 2006). Especially,
through the detection of GRB 030329, the association of long GRBs
with supernovae is firmly established (Hjorth et al. 2003), which
strongly supports the collapsar model as the energy mechanism for
long GRBs (Woosley 1993; MacFadyen \& Woosley 1999). Theoretically,
the collapse of a massive star will most likely give birth to a
black hole, surrounded by a temporal accretion disk. It is a common
sense that the accretion system will produce double-sided jets
(MacFadyen \& Woosley 1999; Aloy et al. 2000; Rhoads 1999;
M\'esz\'aros 2002). The GRB can be observed only when our line of
sight is right on one of the two jets. The collimation of GRB ejecta
can be tested observationally, through various beaming effects, such
as the achromatic break in GRB afterglow light curves (Sari et al.
1999; Liang et al. 2008), the polarization in both the main burst
phase and the afterglow phase (Lazzati 2006), the predicted
existence of orphan afterglows (Rhoads 1997; Huang et al. 2002;
Granot \& Loeb 2003), and the energy crisis already noted in some
GRBs (Frail et al. 2001). In fact, more and more observational
evidences have been accumulated today, supporting the idea that many
GRB ejecta might be highly collimated.

Current studies on the beaming effects are mostly concentrated on
the emission from the forward jet, i.e., the jet moving toward the
observer. The emission from the receding jet (or the counter jet) is
generally omitted. It is interesting to note that this ingredient
recently has been studied by a few authors (Granot \& Loeb 2003; Li
\& Song 2004). By some simple analytical derivations, Li \& Song
(2004) argued that the emission from the receding jet can be
detected in a few cases in the non-relativistic phase of GRB
afterglows. However, previous studies did not consider some
important effects, such as the action of the equal arrival time
surface (EATS). Recently, Zhang \& MacFadyen (2009) presented a
two-dimensional simulation of GRB outflow. The emission from the
receding jet has also been included in their calculations, but they
did not investigate the effects of various parameters on the
receding jet component.

In this paper, we will present our detailed numerical investigation
on the emission from the receding jet of GRB in the deep Newtonian
stage. Although the GRB jet may be complicatedly structured
(M\'esz\'aros et al. 1998; Kumar \& Granot 2003; Huang et al. 2004), and the
circum-burst environment may be wind medium and even associated with
some complex density variations (M\'esz\'aros et al. 1998; Chevalier
\& Li 2000; Gou et al. 2001; Wu et al. 2004), here we will only
consider the simplest situation, i.e, the homogeneous double-sided
jets expanding into a homogeneous interstellar medium, which is
favored by some recent fits (Huang et al. 2000a; Yost et al. 2003).

The structure of our paper is organized as follows.  \S 2 is mainly
a review of the dynamics and radiation model we used in our
calculations. In \S 3 we present the numerical results, together
with our tentative explanations. \S 4 is our conclusion and
discussion.

\section{Model Description}

In the afterglow phase, the GRB ejecta expands into the interstellar
medium (ISM) and is decelerated continuously, giving rise to a
strong external shock. The swept-up electrons are accelerated by the
blastwave, producing the afterglow mainly through synchrotron
radiation. In radio bands, the shell is no longer optically thin, so
that the synchrotron self-absorption should be considered. In our
study, we will use the simplified dynamical equations suggested
by Huang et al. (1999, 2000b), which is consistent with the 
self-similar solution of Blandford \& McKee (1976) in the 
ultra-relativistic phase, and is consistent with the Sedov 
solution (Sedov 1969) in the non-relativistic phase. The beaming 
effects (Rhoads 1997, 1999) can also be conveniently simulated 
in this way.  Here, for
completeness, we first describe the dynamics and the radiation
process briefly.

\subsection{Hydrodynamical Evolution}

In our description, $t$ is the photon arrival time measured in the
lab frame; $R$ is the radial coordinate measured in the burst frame
relative to the initiation point; $m$ is the rest mass of the
swept-up medium; $\theta$ is the half-opening angle of the ejecta;
$\gamma$ is the bulk Lorentz factor of the moving material; $p$ is
the electron distribution index which is typically between 2 and 3;
$n$ is the number density of ISM; $\xi_{\rm e}$ and $\xi^2_{\rm B}$
are the energy equipartition factors for electrons and the comoving
magnetic field. We further denote the initial values of the rest
mass, the isotropic equivalent energy, the Lorentz factor and the
half-opening angle of the ejecta as $M_{\rm ej}, E_{\rm 0,iso},
\gamma_0, \theta_{\rm j}$, respectively.

The overall dynamical evolution of the GRB ejecta can be depicted by
\begin{eqnarray}
\frac{\ud R}{\ud t} & = & \beta c\gamma \left( {\gamma  + \sqrt {\gamma ^2 - 1} } \right), \\
\frac{\ud m}{\ud R} & = & 2\pi \left( {1 - \cos \theta } \right)R^2nm_{\rm p} ,\\
\frac{\ud \theta}{\ud t} & = & \frac{c_ {\rm s}}{R} \left( {\gamma + \sqrt {\gamma ^2 - 1} } \right), \\
\frac{\ud \gamma }{\ud m} & = & - \frac{\gamma ^2  - 1}{M_{\rm ej} +
\varepsilon m + 2(1 - \varepsilon )\gamma m},
\end{eqnarray}
where $\beta = \sqrt {1- 1/\gamma ^2}$, $c$ is the speed of light,
and $c_{\rm s}$ is the comoving sound speed, which can be calculated
by $c_ {\rm s}^2 = \hat \gamma (\hat \gamma - 1)(\gamma - 1) c^2 /
\left[ 1 + \hat \gamma (\gamma - 1) \right] $ with $\hat \gamma
\approx (4\gamma + 1)/({3\gamma })$ being a reasonable approximation
for the adiabatic index. In Equation (4), $\varepsilon$ is the
radiative efficiency. In the extreme case, $\varepsilon = 0$ means
adiabatic condition and $\varepsilon = 1 $ refers to highly
radiative situation. Note that in realistic case, $\varepsilon$
should evolve gradually from 1 to 0, in about several hours.

Equations (1) --- (4) is a convenient description of GRB afterglow
dynamics that is applicable in both the initial ultra-relativistic
phase and the late Newtonian phase.

\subsection{Radiation Process}

Basically, we assume that the shock-accelerated electrons follow a
power-law distribution according to their energies, $\ud N'_{\rm e}
/ {\ud \gamma_{\rm e}} \propto \gamma_{\rm e}^{- p}$,
 However, to ensure that the calculation in the deep Newtonian phase is correct, we
need to modify the basic distribution function as $\ud N'_{\rm e} /
{\ud \gamma_{\rm e}} \propto \left( \gamma_{\rm e} -1 \right)^{- p}$
(Huang \& Cheng 2003). The minimum and maximum Lorentz factors of
electrons can be calculated as $\gamma_{\rm e,\min} = \xi _e (\gamma
- 1) m_{\rm p} (p - 2)/[ m_{\rm e} (p - 1) ] + 1$ and $\gamma_{\rm
e,\max} = \sqrt { 6\pi e / \left( \sigma_{\rm T} B' \right) }
\approx  10^8 (B'/ 1 {\rm G})^{-1/2}$, where $B'$ is the comoving
magnetic field strength, $m_{\rm p}$ and $m_{\rm e}$ are masses of
proton and electron, respectively. As usual, we assume that the
energy ratio of magnetic field with respect to internal energy is
$\xi^2_{\rm B}$, so that the energy density of magnetic field is
$B'^2/(8\pi) = \xi^2_{\rm B} \left( \hat \gamma  - 1 \right)^{-1} (
\hat \gamma \gamma + 1)(\gamma  - 1) n m_{\rm p} c^2 $.

The cooling of electrons due to synchrotron radiation will lead to a
steep distribution function above a critical Lorentz factor,
$\gamma_{\rm c}$. The expression for $\gamma_{\rm c}$ can be derived
as $\gamma_{\rm c} = 6\pi m_{\rm e} c / \left( \sigma_{\rm T} \gamma
B'^2 t \right)$, where $\sigma_{\rm T}$ is the Thompson scattering
cross section (Sari et al. 1998). Considering all the above
ingredients, we finally use the following electron distribution
function in our calculations (Huang \& Cheng 2003):  \\
1. $\gamma_{\rm c} \le \gamma_{\rm e,\min} $,
\begin{equation}
\frac{\ud N'_{\rm e}}{\ud \gamma _{\rm e} } \propto \left\{
\begin{array}{l}
 \left( \gamma _{\rm e} - 1 \right)^{- 2}   \hspace{0.8cm} (\gamma _{\rm c}  \le \gamma _{\rm e}  < \gamma _{\rm e,\min } ), \\
 \left( \gamma _{\rm e} - 1 \right)^{ -(p + 1)}  \hspace{0.335cm}  (\gamma _{\rm e,\min }  \le \gamma _{\rm e}  \le \gamma _{\rm e,\max } ); \\
 \end{array} \right.
\end{equation}
2. $\gamma_{\rm e,\min} < \gamma _{\rm c} \le \gamma _{\rm e,\max}$,
\begin{equation}
\frac{\ud N'_{\rm e}}{\ud \gamma _{\rm e} } \propto \left\{
\begin{array}{l}
 \left( \gamma _{\rm e} - 1 \right)^{ - p} \hspace{0.78cm}  (\gamma _{\rm e,\min }  \le \gamma _{\rm e}  \le \gamma _{\rm c} ), \\
 \left( \gamma _{\rm e} - 1 \right)^{ -(p + 1)} \hspace{0.34cm}  (\gamma _{\rm c}  < \gamma _{\rm e}  \le \gamma _{\rm e,\max } ); \\
 \end{array} \right.
\end{equation}
3. $\gamma _{\rm c}  > \gamma _{\rm e,\max}$,
\begin{equation}
\frac{\ud N'_{\rm e}}{\ud \gamma _{\rm e}} \propto \left( \gamma
_{\rm e} - 1 \right)^{- p} \hspace{0.55cm}  (\gamma _{\rm e,\min }
\le \gamma _{\rm e}  \le \gamma _{\rm e,\max } ).
\end{equation}

In the comoving frame, the synchrotron radiation power at $\nu'$ is
(Rybicki \& Lightman 1979)
\begin{equation}
P'(\nu') = \frac{\sqrt 3 e^3 B'}{m_{\rm e} c^2 } \int_{\min \left(
\gamma_{\rm e,\min}\ ,\ \gamma_{\rm c} \right)}^{\gamma_{\rm e,\max}
} {\left( \frac{\ud N'_{\rm e}} {\ud \gamma_{\rm e}} \right)}\
F\left( \frac{\nu'}{\nu'_{\rm e }} \right) d\gamma_{\rm e},
\end{equation}
with $F(x) = x\int_x^{ + \infty } {K_{5/3}(k) \, \ud k} $ being the
Bessel function and $\nu'_{\rm e} = 3\gamma^2_{\rm e} e B' / \left(
4\pi m_{\rm e} c \right)$ being the characteristic emission
frequency (Shu 1991; Longair 1992). 
To calculate the radio afterglows, we
must consider the synchrotron self-absorption. The optical depth of
synchrotron self-absorption can be obtained as
\begin{equation}
\tau _{\nu '} = \frac{\sqrt 3 e^3 B'}{8 \pi m^2_{\rm e} c^2 \nu '^2
} \int_{\min \left( \gamma_{\rm e,\min}\ ,\ \gamma_{\rm c}
\right)}^{\gamma_{\rm e,\max } } {(q + 2)\left( \frac{\ud n'_{\rm e}
}{\ud \gamma_{\rm e} } \right)\frac{1}{\gamma_{\rm e} }}\ F\left(
\frac {\nu'}{\nu'_{\rm e} } \right)d\gamma_{\rm e},
\end{equation}
where $\ud n'_{\rm e} /\ud \gamma_{\rm e}$ denotes the column
density distribution of electrons measured in the comoving frame
on the line of sight; $q$ is the electron power-law
distribution index which varies from $2$ to $p+1$ for fast-cooling
and from $p$ to $p+1$ for slow-cooling. The synchrotron
self-absorption will affect the radiation by a reduction-factor
(Waxman et al. 1998)
\begin{equation}
f(\tau) = \frac {1- e^{-\tau _{\nu '}}} {\tau _{\nu '}}.
\end{equation}

Let us define the Doppler-factor as $D = \left[ \gamma \left( 1 -
\beta \mu \right) \right]^{- 1} $ (M\'esz\'aros 2006), where $\mu
=\cos \Theta$ and $\Theta $ is the angle between the velocity of the
emitting material and the line of sight. Also we denote the viewing
angle as $\theta_{\rm obs}$. Then the observed frequency is $ \nu =
D \nu' /(1 + z)$, and the observed flux density from a point-like
source is
\begin{equation}
F_\nu = \frac{(1 + z) D^3 }{4\pi d_{\rm L}^2 } f(\tau) P'\left[ (1 +
z)D^{-1} \nu \right],
\end{equation}
where $d_{\rm L}$ is the luminosity distance. Finally, we can
integrate the flux density over the EATS (Waxman 1997; Sari 1998) 
determined by 
\begin{equation}
t_ {\rm obs} = (1+z) \int \frac{\ud R}{\beta \gamma c D} \equiv
\rm{const}.
\end{equation}

\section{Numerical Results}

In this section, we present our numerical results concerning the
emission from the receding jet. First of all, for simplicity, we
assume that the twin jets have the same characteristics, i.e., the
same initial energy, opening angle, initial Lorentz factor, and the
circum-burst ISM density. We also assume that the microphysics shock
parameters ($p$, $\xi_{\rm e}$, $\xi^2_{\rm B}$) are the same for
the receding and forward blastwaves. For convenience, we define a
set of parameter values as the ``standard'' condition: $n=1 / {\rm
cm}^{3}$, $E_{\rm 0,iso} =10^{53} {\rm ergs}$, $\theta_{\rm j} =0.1
$, $\varepsilon = 0 $, $\xi_{\rm e} = 0.1$, $\xi^2_{\rm B}=0.01$,
$p=2.5$, $\theta_{\rm obs}=0$, and $\gamma_0 = 300$. These values
are typical in the study of GRB afterglows. For redshift, we adopt
the value of $z=0.1$ (which corresponds to $d_{\rm L}= 454$ Mpc
according to the popular cosmology model, Wright 2006).

Firstly we illustrate the evolution of the Lorentz factors of the
twin jets in Fig.~1. Note that the X-axis is observers' time. For
the observer, the dynamical evolution of the receding jet is quite
different from that of the forward jet, especially in the
relativistic phase. We see that in a rather long time ($t \sim 50$
d), $\gamma$ of the receding jet remains almost constant. This is
due to the time delay induced by the long distance between the twin
jets. It also implies that the emission from the receding jet will
be very weak in this period, since it is highly beamed backwardly.
At the observers' time of $t \sim 340$ d, the Lorentz factor of the
receding jet is still more than 10, while the forward jet's Lorentz
factor has already decreased to less than 1.1 .

In Fig.~2, we show some examples of the equal arrival time surfaces
(EATSes) at three moments. As expected, at any particular moment,
the typical radius of the surface is much larger for the forward jet
branch as compared with that for the receding jet branch. Also, we
notice that the curvature of the two branches is quite different.
Generally, the EATS is much flatter on the receding jet. Another
interesting feature is that the area of the EATS on the forward
branch is much larger than that of the corresponding receding
branch.

Fig.~3 shows the radio and optical afterglow light curves under the
``standard'' condition (thick lines). Here, the thick dotted line
corresponds to emission from the forward jet, the thick dashed line
corresponds to emission from the receding jet, and the thick sold
line is the total light curve. Under the ``standard'' condition, for
the forward jet, the afterglow light curve (the dotted line) becomes
slightly flattened in the non-relativistic phase. It is consistent
with previous results in the deep Newtonian phase (Huang \&
Cheng 2003). Also it can be seen that the receding jet really can
contribute a significant portion in the total emission at very late
stage. The role played by the receding jet is reasonably more
important in the radio band than in the optical band. However, the
dashed component is generally not very strong, so that it can only
lead to a plateau in the total light curve, but not an obvious
rebrightening or a marked peak as expected by Li \& Song (2004).
Interestingly, our result is consistent with the simulation of Zhang
\& Macfadyen (2009). We believe that the discrepancy between our
numerical result and Li \& Song's analytical result mainly comes
from the effect of the EATS. Below, we will give some detailed
analyses on this point. Additionally, it should be noted that in the
radio band, the peak flux of the receding component is about 4
orders of magnitude weaker than that of the forward component. It
essentially means that the receding component is very weak, and is
very difficult to detect. In the optical band, the condition is even
more awkward. The peak flux of the receding component is about 23
magnitudes dimmer than that of the forward component in R band. Even
comparing with the flux of the forward jet at the jet break time, it
is still 16 --- 17 magnitudes weaker. So, in optical band, it is
even much more difficult to observe the receding jet component.

According to Li \& Song (2004), the time when the receding jet
becomes notably visible ($t_{\rm NR}^{\rm RJ}$) is relevant to the
time when the forward jet enters the non-relativistic phase ($t_{\rm
NR}$), i.e.,
\begin{equation}
t_{\rm NR}^{\rm RJ} = t_{\rm NR} + \frac {2r_{\rm NR}}{c},
\end{equation}
where $r_{\rm NR}$ is the radius of the forward jet at $t_{\rm NR}$.
In the standard frame work (Blandford \& Mckee 1976; Rhoads 1999),
the sphere-like phase of a highly collimated GRB ejecta ends at the
so called jet break time determined by $\gamma _ {\rm j} = 1/
\theta_{\rm j}$, with the shock radius being $r_{\rm j}= \left( 3
E_{\rm 0,iso} \theta_{\rm j}^2 / \left[ {4\pi n m_{\rm p}
c^2}\right] \right)^ {1/3}$. After the sphere-like phase, the jet
spreads laterally at the co-moving sound speed  $c_{\rm s}$ so that
we have  $\gamma \propto t^{-1/2}$ and $r_{\rm NR} \approx r_{\rm
j}$ (Rhoads 1999).  Then finally we obtain
\begin{equation}
 t_{\rm NR} = \frac{1}{2c} \left( \frac{3 E_{\rm 0,iso} \theta_{\rm j}^2}{4\pi n m_{\rm p} c^2} \right) ^{1/3},
 \\ \qquad t_{\rm NR}^{\rm RJ} = 5t_{\rm NR}.
\end{equation}

Adopting the standard values of our parameters, Equation (14) yields
$t_{\rm NR} \approx 104$ d and $t_{\rm NR}^{\rm RJ} \approx 520$ d.
After correcting for the cosmological time dilation ($z=0.1$), we
get the corresponding observers' time of $t_1=(1+z)\ t_{\rm NR}
\approx 114$ d and $t_3=(1+z)\ t_{\rm NR}^{\rm RJ} \approx 572$ d.
In fact, in Fig.~2, the EATSes for these two moments have been
displayed. So, according to Li \& Song's suggestion, the
contribution from the receding jet should peak at $t_3 \approx 572$
d. In our Fig.~3, for the ``standard'' condition, the peak is
postponed to $t_{\rm peak} \sim 1140 $ d for 8.46 GHz, and to
$t_{\rm peak}\sim 1700$ d in R band. So, the EATS effect and the
deceleration of the external shock can lead to some subtle
difference between the analytical results and the numerical results.
Actually, Zhang \& MacFadyen's numerical results have clearly shown
that the observers' time does not equal to the burst frame time at
$t_{\rm NR}$ (Zhang \& MacFadyen 2009). Unfortunately, in previous
analysises it is usually assumed that these two times are equal.


Another reason that suppresses the rebrightening of the receding jet
is as follows. According to Li \& Song's analysis, at the observers'
time $t_3$, the receding jet should be at the radius of $r_{\rm
NR}$. However, from our Fig.~2, we see that the typical radius of
the EATS at $t_3$ on the receding jet is much smaller than the
radius of the forward jet at $t_1$. The reason is again due to the
EATS effect. It means that the receding jet still does not
decelerate enough at $t_3$ (actually, the bulk Lorentz factor is
still 3.95), and its emission is still mainly directed forwardly
(not backwardly toward the observer). Additionally, Fig.~2 shows
clearly that the area of the receding jet at $t_3$ (corresponding to
$t_{\rm NR}^{\rm RJ}$ ) is much smaller than that of the forward jet
at $t_1$  (corresponding to $t_{\rm NR}$). So, the number of
electrons involved in the radiation process is typically much
smaller on the receding jet at $t_{\rm NR}^{\rm RJ}$, as compared to
that on the forward jet at $t_{\rm NR}$. Due to the above reasons,
the contribution from the receding jet is naturally much weaker than
that deduced from $L_{\rm \nu}^{\rm RJ}(t) \approx L_{\rm
\nu}(t-4t_{\rm NR}),\ (t\geq t_{\rm NR}^{\rm RJ}) $ (Equation (7) in
Li \& Song (2004) ).

However, although the receding jet emission is generally very weak
in our ``standard'' condition, we guess that in some special cases
it still can be enhanced. Obviously, a denser environment will help
to decelerate the jet more quickly, thus lead to a smaller $t_{\rm
peak}$ and a higher intensity.
In Fig.~3, we have also plotted in thin lines our numerical results
for a double-sided jet that locates in a dense circum-burst medium
($n=1000 / {\rm cm}^{3}$). 
Note that other parameters involved here are the same as the
``standard'' case. Encouragingly, in Fig.~3(a) we see that the peak
time of the receding jet can be as early as $t_{\rm peak} \sim 150$
d, with the peak flux as large as a few mJy in radio band (i.e.,
only several times less than the peak level of the forward jet). In
Fig.~3(b), the optical contribution from the receding jet is still
very weak, with the peak flux being about 28$^m$.

In Fig.~4, we plot the afterglow light curves in more radio and
optical/infrared bands. Generally speaking, $t_{\rm peak}$ is about
1140 d in radio bands and is about 1700 d in optical bands. Such a
difference in the peak time is insignificant, considering that the
frequency difference between radio and optical wavelengths is really
huge. We notice that $t_{\rm peak}$ almost remains the same from
radio to X-ray bands in Fig.~7 of Zhang \& MacFadyen (2009). Thus
our results are roughly consistent with Zhang \& MacFadyen's.
Another interesting conclusion that can be drawn from our Figs. 3
and 4 is that at lower frequency, the relative intensity of the
receding jet component (its peak flux), as compared with the peak of
the forward jet component, becomes stronger. Such a tendency can
also be roughly seen in Fig.~7 of Zhang \& MacFadyen (2009).

Fig.~5 illustrates the effects of some parameters ($n$, $E_{\rm 0,
iso}$, $\theta_{\rm j}$, and $\varepsilon$) on the receding jet
component in the afterglow light curve. Fig.~5(a) shows that the
circum-burst medium density ($n$) affects the peak time ($t_{\rm
peak}$) of receding jet dramatically. A larger number density
usually leads to a smaller $t_{\rm peak}$. The strength of the
receding jet component is also obviously enhanced.
It again hints that the receding jet component is most likely
detectable in a dense environment. Similarly, the initial kinetic
energy ($E_{\rm 0, iso}$) also affects $t_{\rm peak}$ significantly,
with larger $E_{\rm 0, iso}$ corresponding to a larger $t_{\rm
peak}$ (Fig.~5(b)). The effect of the initial jet opening angle
($\theta_{\rm j}$) on $t_{\rm peak}$ can also be clearly seen in
Fig.~5(c). It should be further noted that the receding jet
component is more marked when the opening angle is smaller. In
Fig.~5(d), we can observe an obvious rebrightening when the
radiation efficiency ($\varepsilon$) is large. However, in realistic
case, $\varepsilon$ is unlikely to be so large. Actually, at such
late stages, the external shock should be adiabatic, so that
$\varepsilon$ should be nearly zero.

In Fig. 5(d), we also plot the radio afterglow light curves for
double-sided jets under some special physical assumptions. The
dash-dotted line is plotted by assuming that both the forward jet
and the receding jet do not experience any lateral expansion. Since
the deceleration of the jets is much slower in this case, we see
that the receding jet component emerges much later and is also much
less obvious as compared with our ``standard'' case. The dotted line
is plotted by assuming a much smaller initial Lorentz factor
($\gamma_0 = 30$), which may correspond to the so called failed GRBs
(Huang et al. 2002). The receding jet component emerges slightly
earlier as compared with the solid line, but its role becomes less
significant correspondingly.

Fig.~6 illustrates the effects of other four parameters ($\xi_{\rm
e}$, $\xi_{\rm B}^2$, $p$, and $\theta_{\rm obs}$) on the receding
jet component. Generally speaking, a larger $\xi_{\rm e}$ and/or
$\xi_{\rm B}^2$ can enhance the receding jet component markedly. On
the other hand, although $p$ has an important influence on the
overall afterglow light curve, its impact on the relative strength
of the receding jet component is not significant. Again, note that
in all the cases, the contribution from the receding jet only
emerges as a plateau, but not as any obvious rebrightening. In
Fig.~6(d), when the observing angle ($\theta_{\rm obs}$) increases,
the forward jet component becomes weaker, while the receding jet
component becomes stronger. It is in good accord with our
expectation (also see Granot \& Loeb 2003). However, the
contribution from the receding jet still generally plays a minor
role in the total afterglow light curve. Additionally, for off-axis
twin jets, the GRB from the forward jet is un-observable, so that
even the afterglow from the forward jet itself (i.e., the orphan
afterglow) is difficult to observe. Note that in Fig.~6(d), when
$\theta_{\rm obs} = \pi / 2$ (i.e., the thick solid line), the
contribution from the receding jet and the forward jet are actually
equal.

Equation (14) tells us that the peak time of the receding component
should be relevant to the 3 parameters of $n$, $E_{\rm 0,iso}$,
$\theta_{\rm j}$; on the other hand, other parameters such as
$\xi_{\rm e}$, $\xi_{\rm B}^2$, $p$ do not affect the peak time.
These tendency can be clearly seen in Figs. ~5 and 6.

In all the above calculations, we have assumed that the conditions
and parameters of the twin jets are the same. However, this may not
be the case for realistic GRBs. The circum-burst environment and the
micro-physics parameters may actually be different for the twin
jets, as that may happen in the two component jet structure (Huang
et al. 2004; Jin et al. 2007; Racusin et al. 2008). In Fig.~7, we
have plotted the overall afterglow light curves by assuming
different parameters for the forward jet and the receding jet. In
each panel of Fig.~7, we first plot a common light curve (the solid
line) by adopting the standard parameter set, but change $\xi_{\rm
e}$ to 0.01 and change $\xi^2_{\rm B}$ to $10^{-4}$. We then
increase the values of $\xi_{\rm e}$, $\xi_{\rm B}^2$, and $n$ for
the receding jet to see their effects on the afterglow light curve.
It is encouraging to see that the emission from the receding jet
really can be greatly enhanced, so that it can manifest as an
obvious rebrightening in the overall light curve. In Fig.~7(a), 7(b)
and 7(d), the peak flux of the rebrightening can be nearly 100 times
larger than the ``background'' level in the best cases. It is
imaginable that in the most favorable cases, when all $\xi_{\rm e}$,
$\xi_{\rm B}^2$ and $n$ are larger for the receding jet at the same
time, the rebrightening will be even more remarkable. However, note
that the contrary condition may also exist in realistic GRBs, i.e.,
these parameters may also be smaller for the receding jet. Then the
emission from the receding jet will be completely unnoticeable.

\section{Conclusion and Discussion}

We have studied the emission of the receding jet numerically. The
effect of the EATS is included in our calculations. Clearly, this
effect plays an important role in the process. It is found that the
contribution from the receding jet is generally quite weak. In most
cases, it only manifests as a short plateau in the overall afterglow
light curve, but not a marked rebrightening. The flux density of the
plateau is usually much less than 100 $\mu$Jy in radio bands even at
a small redshift of $z=0.1$ . If we place the GRB at a more typical
redshift of $z=1$, then the flux density of the plateau will be less
than 0.1 $\mu$Jy at 8.46 GHz. We noticed that the observed radio
afterglow emission is generally on the level of 0.1 --- 1 mJy at
about the peak time. After several months, the radio afterglow
usually decreases to a very low level, and is submersed by the
emission from the host galaxy, whose strength can be 40 --- 70
$\mu$Jy (Berger et al. 2001). Additionally, the error bar of radio
observations is usually $\sim $ 30 --- 50 $\mu$Jy at very late
stages (Frail et al. 2003). Thus the contribution from the receding
jet, i.e. the plateau, is actually very difficult to detect
currently, especially for those GRBs at $z \sim 1$. Our results are
consistent with a recent observational report by van der Horst et
al. (2008), who failed to detect any clear clues of the receding jet
emission.

However, as shown in our Fig.~7, if the micro-physics parameters of
the receding jet were different from the forward jet, or if the
receding jet were in a much denser environment, then it is still
possible that the contribution from the receding jet can be greatly
enhanced. For example, if $\xi_{\rm e}$ and/or $\xi_{\rm B}^2$ of
the receding jet is much larger than that of the forward jet, then
the receding jet can really manifest as an obvious rebrightening.

Also, our Fig.~5(a) shows that a dense circum-burst environment can
suppress the emission of the forward jet, and enhance the
contribution from the receding jet. If the GRB occurs in a very
dense molecular cloud with $n > 10^3 / {\rm cm}^{3}$ (Dai \& Lu
1999), the contribution from the receding jet may be much easier to
detect. Additionally, if the GRB is very near to us at the same
time, then the possibility of successfully detecting the receding
jet is very high (see the thin lines in Fig.~3(a)).

In short, we believe that the effort of trying to search for the
afterglow contribution from the receding jet is still meaningful. If
observed, it would provide useful clues to study the circum-burst
environment and the micro-physics of external shocks. We suggest
that nearby GRBs (with redshift $z \leq 0.1$) should be good
candidates for such studies.

\begin{acknowledgements}
We would like to thank the anonymous referee for constructive
suggestions that lead to an overall improvement of this study.
We also thank Z. Li for stimulating discussion. This research was
supported by the National Natural Science Foundation of China (grant
10625313), and by the National Basic Research Program of China
(grant 2009CB824800). Xin Wang is also supported by 2008' National
Undergraduate Innovation Program of China (grant 081028441).
\end{acknowledgements}

\clearpage

\begin{figure*}
   \begin{center}
   \includegraphics[width=0.7\textwidth]{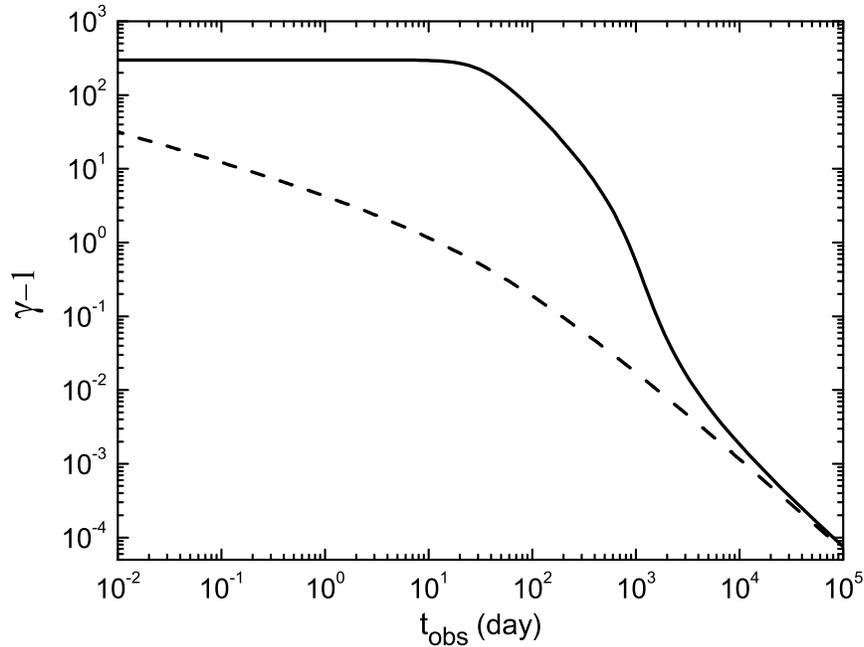}
   \par\end{center}
   \caption{The evolution of the Lorentz factors of the twin jets. The solid line corresponds
   to the receding jet and the dashed line is plotted for
   the forward jet. The twin jets are in ``standard'' condition as defined in Sect.~3.
   The observers' time has been corrected for the cosmological effect ($z=0.1$).}
   \label{fig1}
\end{figure*}

\begin{figure*}
   \begin{center}
   \includegraphics[width=0.8\textwidth]{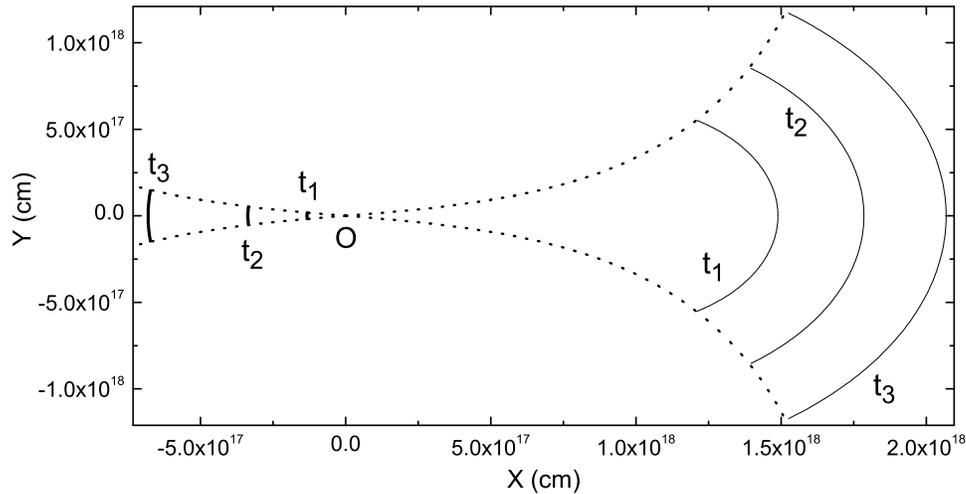}
   \par\end{center}
   \caption{Schematic illustration of the EATSes at three
    moments, $t_1 \approx 114$ d, $t_2 \approx 286$ d and $t_3 \approx 572$
    d. In this calculation, we have used the ``standard'' parameter set as defined
    in Sect.~3. ``O'' is the position of the central engine, and the observer is
    on the far right side with Y=0. The dotted lines indicate the jet boundary.
    For the receding jet, the EATSes are plotted in thick solid lines, while for
    the forward jet the surfaces are plotted in thin solid lines. Note that on
    the forward jet branches, the bulk Lorentz factors of the material at the
    peak of the EATSes  are 1.17, 1.07, and 1.03 for $t_1$, $t_2$, and $t_3$,
    respectively. On the receding jet branches, the bulk Lorentz factors of the
    material at the peak of the EATSes are 56.07, 11.79, and 3.95 for $t_1$,
    $t_2$, and $t_3$, respectively.}
    \label{fig2}
\end{figure*}

\begin{figure*}
   \begin{center}
   \includegraphics[width=0.35\paperwidth]{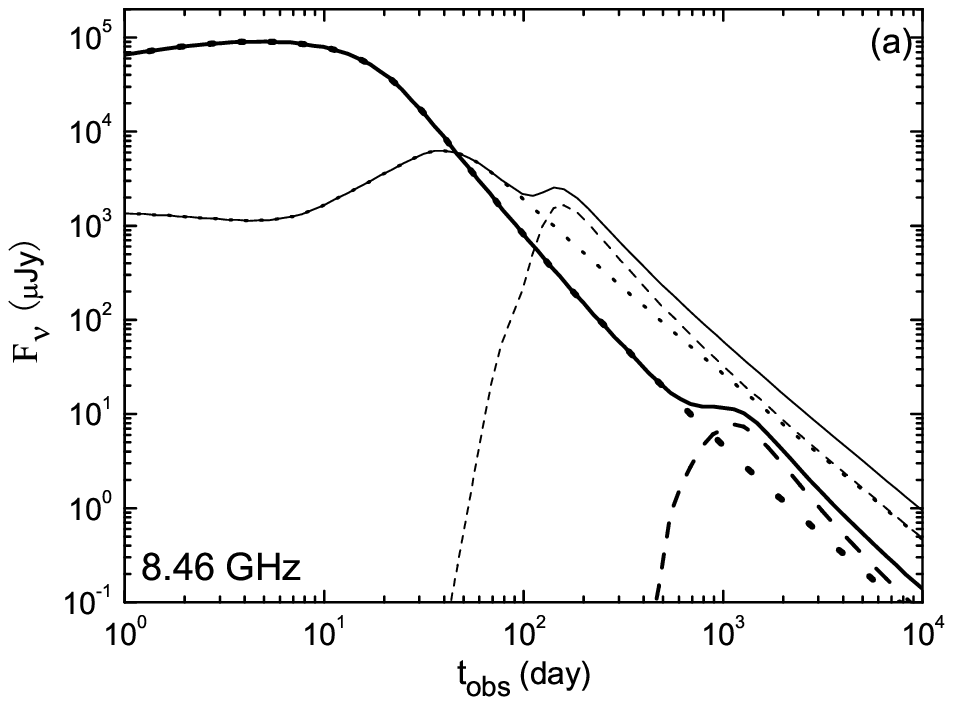}
   \includegraphics[width=0.35\paperwidth]{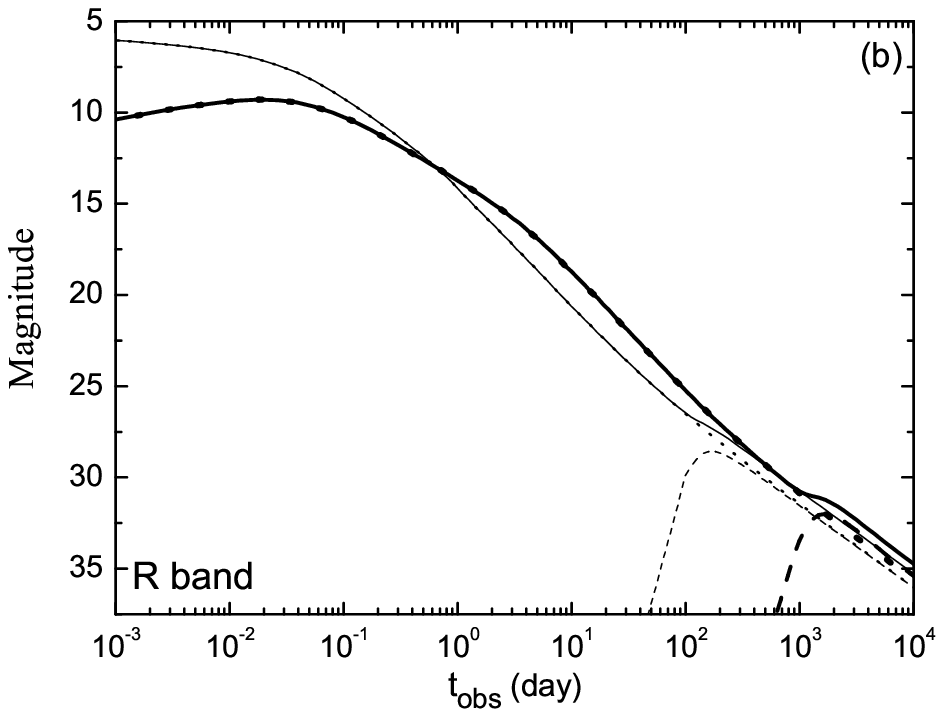}
   \par\end{center}
   \caption{8.46 GHz radio afterglow (a) and R-band optical afterglow (b) from
   the forward jet and the receding jet. The thick lines are plotted
   for a ``standard'' double-sided jet as defined in Sect.~3. The thin lines are plotted
   for the double-sided jet with only one parameter altered as compared with the ``standard''
   condition, i.e. $n=1000 / {\rm cm}^{3}$. In each group, the dotted line
   reflects the emission from the forward jet, the dashed line
   reflects the contribution from the receding jet, and the solid line is the total light curve.}
   \label{fig3}
\end{figure*}

\begin{figure*}
   \begin{center}
   \includegraphics[width=0.35\paperwidth]{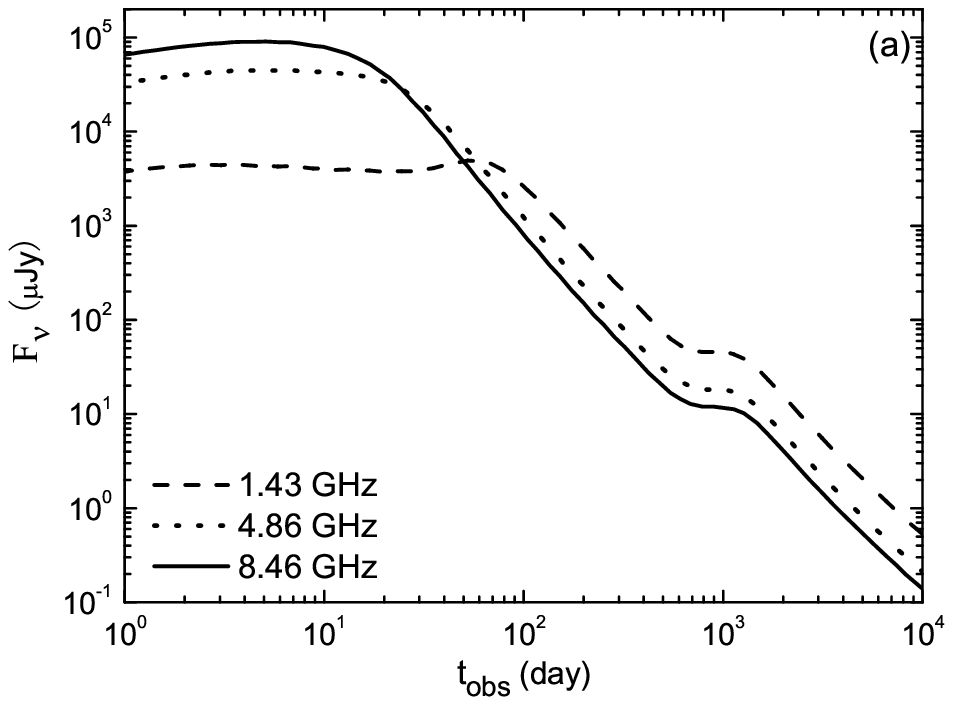}
   \includegraphics[width=0.35\paperwidth]{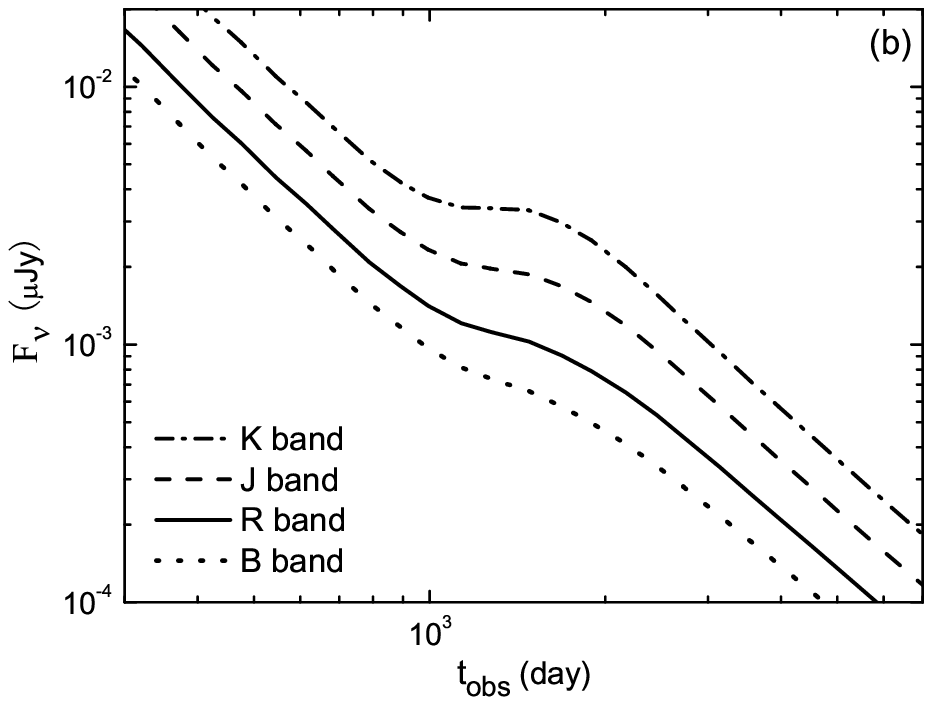}
   \par\end{center}
   \caption{Multiwavelength afterglow light curves of a double-sided jet.
   Radio afterglows are illustrated in panel (a), and optical/IR afterglows
   are plotted in panel (b). In this calculation, we have used the ``standard''
   parameter set as defined in Sect.~3.}
   \label{fig4}
\end{figure*}

\begin{figure*}
   \begin{center}
   \includegraphics[width=0.35\paperwidth]{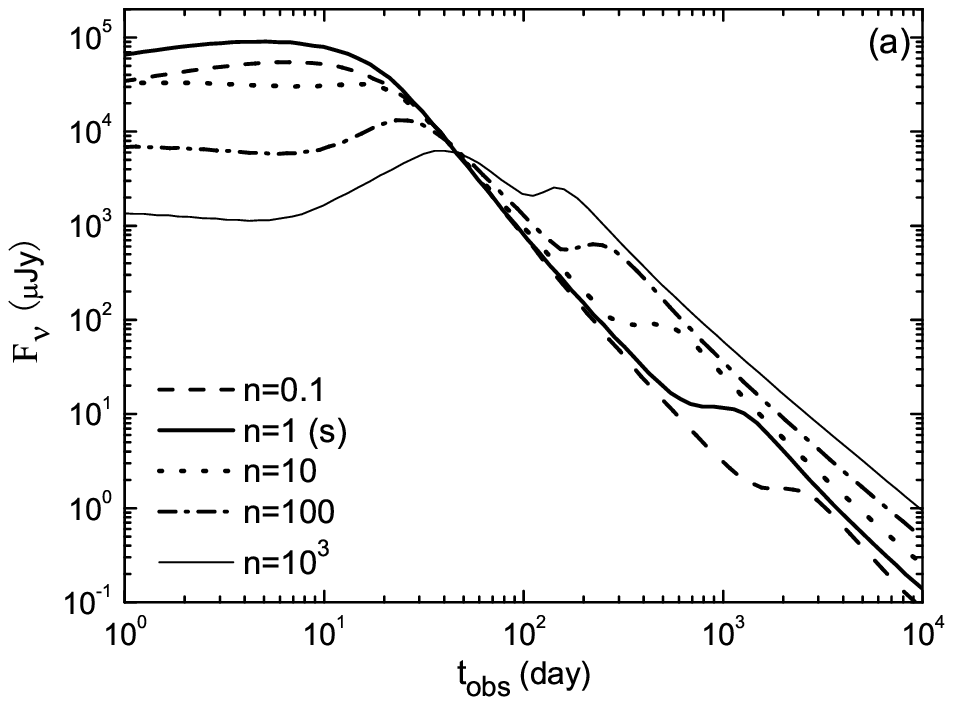} \includegraphics[width=0.35\paperwidth]{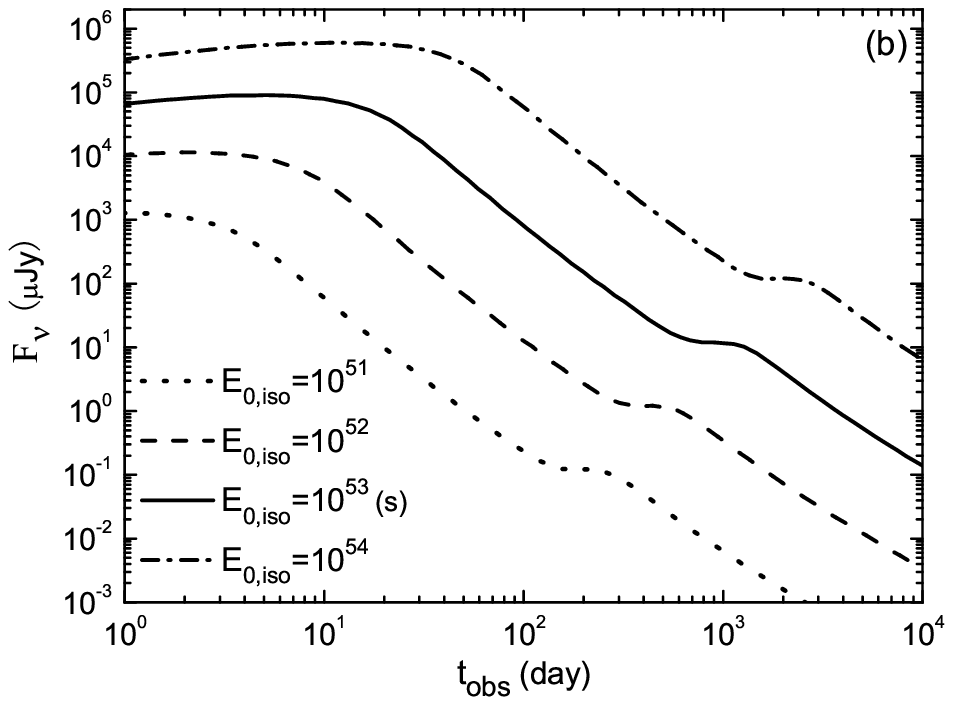}
   \includegraphics[width=0.35\paperwidth]{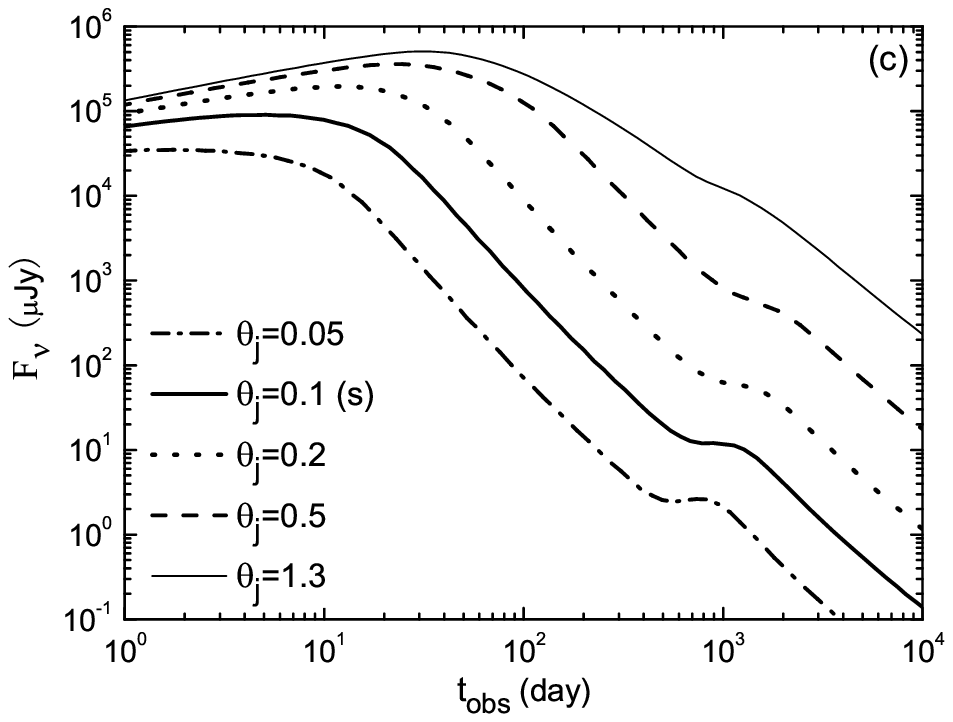} \includegraphics[width=0.35\paperwidth]{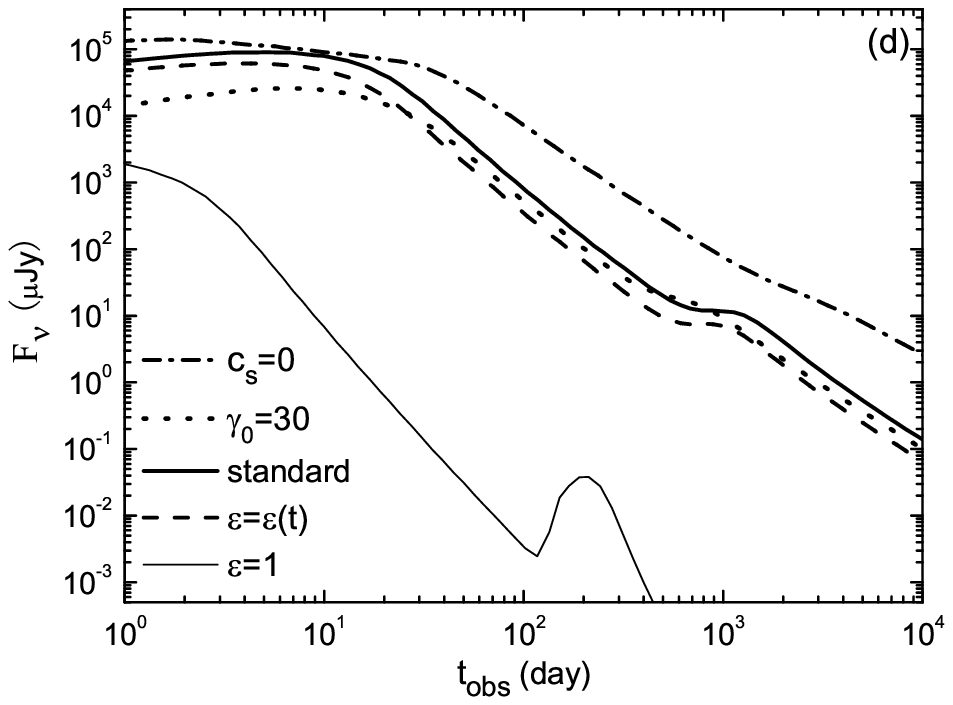}
   \par\end{center}
   \caption{The effects of various parameters ($n$, $E_{\rm 0,iso}$, $\theta_{\rm j}$,
    and $\varepsilon$) on the 8.46 GHz radio afterglow light curves of double-sided jets.
    In each panel,  ``(s)'' corresponds to the ``standard'' condition as defined in
    Sect.~3, and other lines are drawn with only one certain parameter altered or one condition
    changed. In panel (d), the dash-dotted line is plotted for a double-sided
    jet without lateral expansion; and the dotted line is plotted for a double-sided jet with
    a low initial Lorentz factor ($\gamma_{\rm 0} = 30$), which
    may correspond to the so called failed GRBs.}
   \label{fig5}
\end{figure*}

\begin{figure*}
   \begin{center}
   \includegraphics[width=0.35\paperwidth]{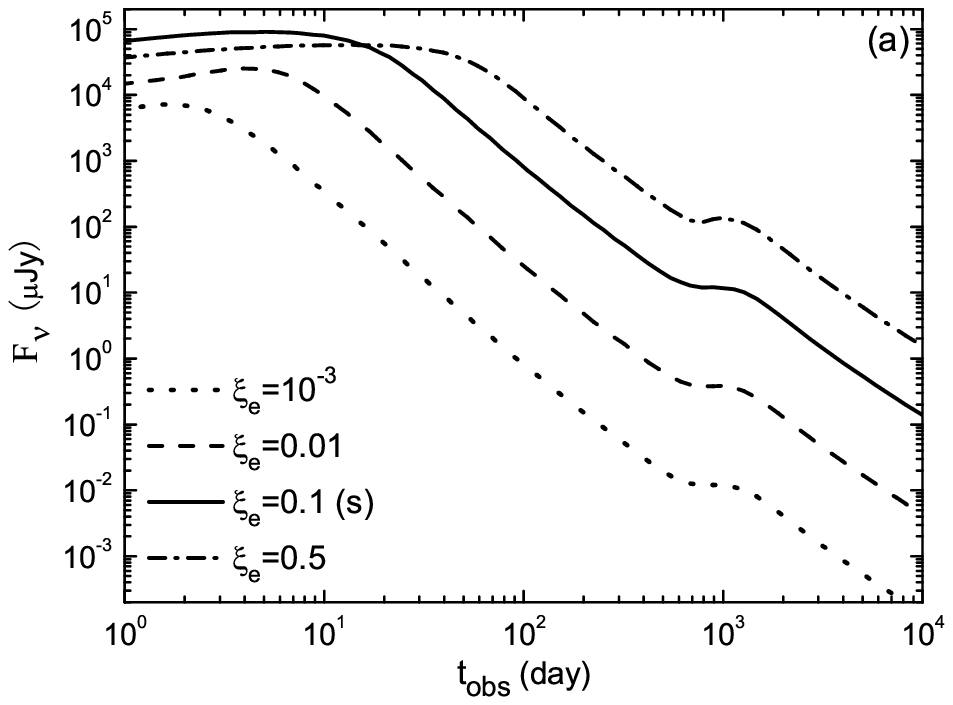} \includegraphics[width=0.35\paperwidth]{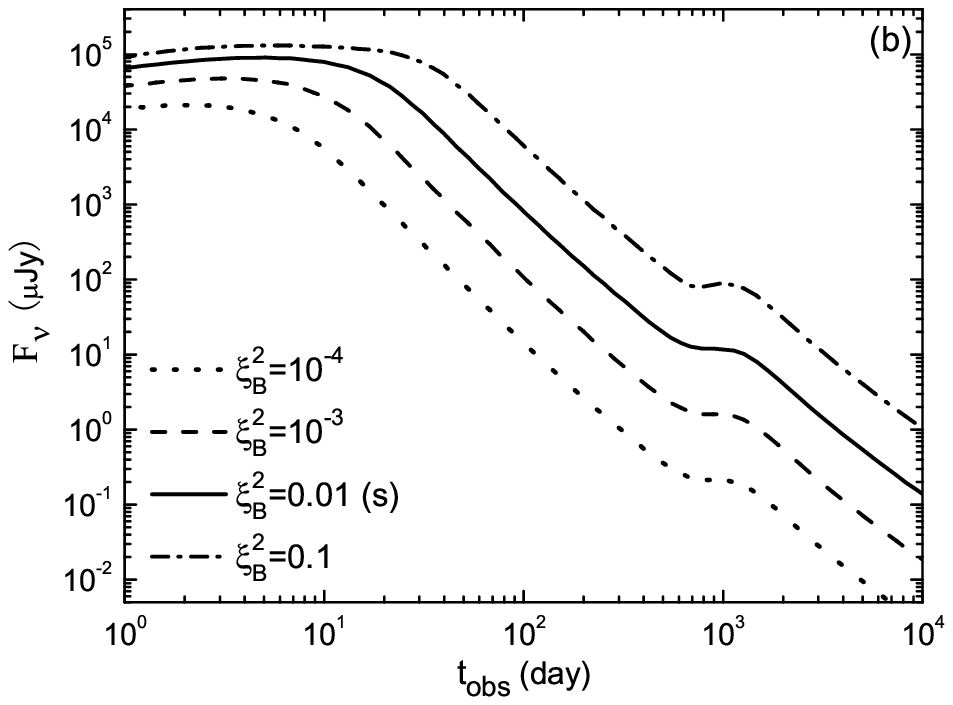}
   \includegraphics[width=0.35\paperwidth]{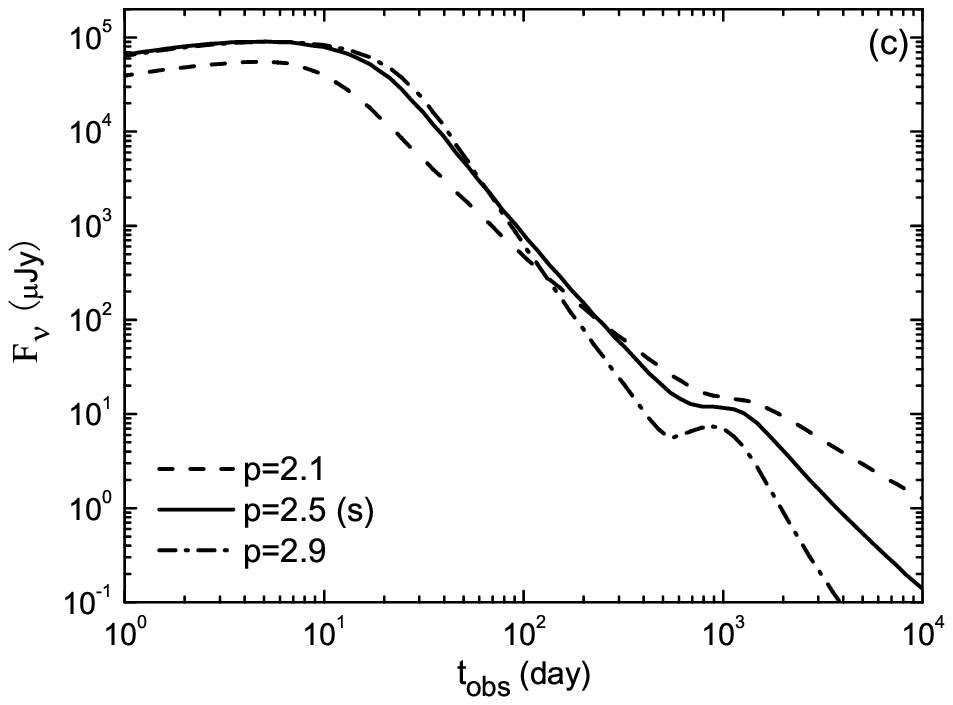} \includegraphics[width=0.35\paperwidth]{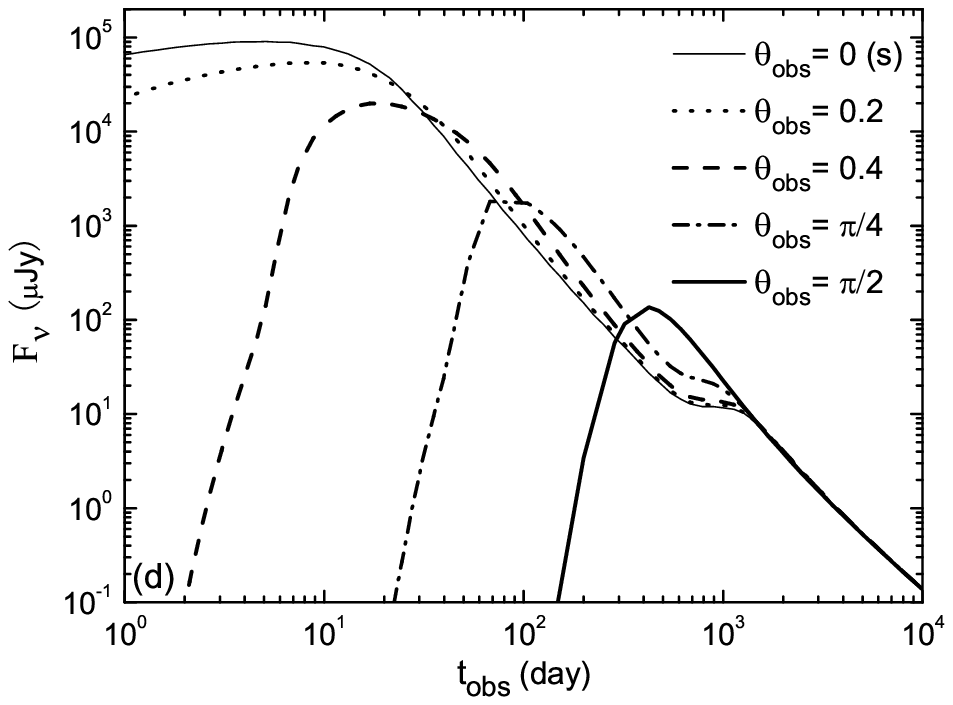}
   \par\end{center}
   \caption{The effects of various parameters ($\xi_{\rm e}$, $\xi_{\rm B}^2$, $p$,
    and $\theta_{\rm obs}$) on the 8.46 GHz radio afterglow light curves of double-sided jets.
    In each panel,  ``(s)'' corresponds to the ``standard'' condition as defined in
    Sect.~3, and other lines are drawn with only one certain parameter altered.}
   \label{fig6}
\end{figure*}

\begin{figure*}
   \begin{center}
   \includegraphics[width=0.35\paperwidth]{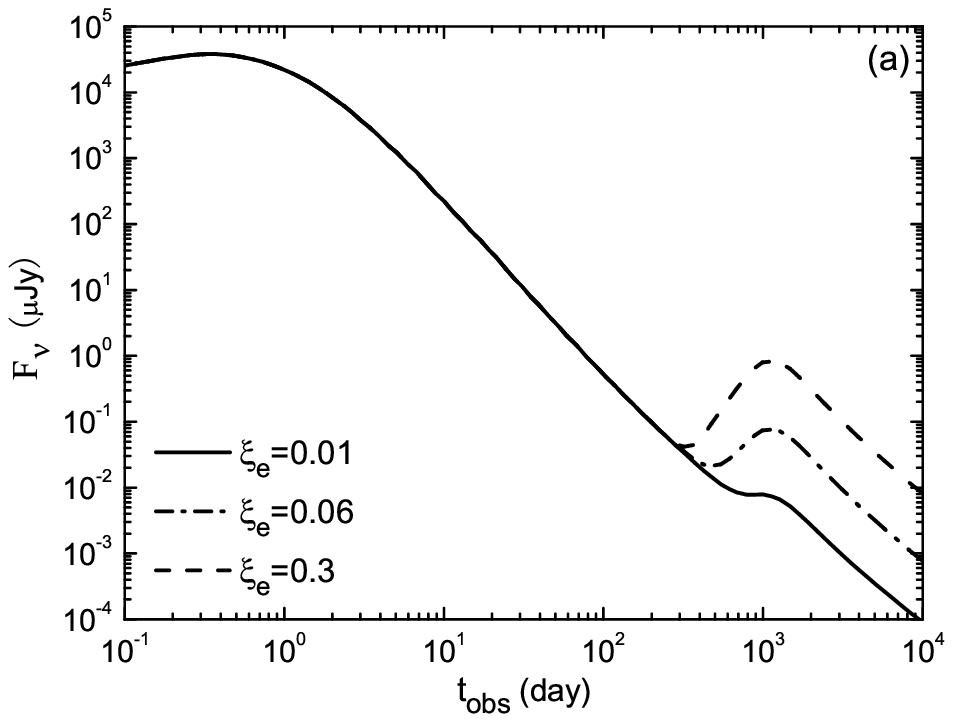} \includegraphics[width=0.35\paperwidth]{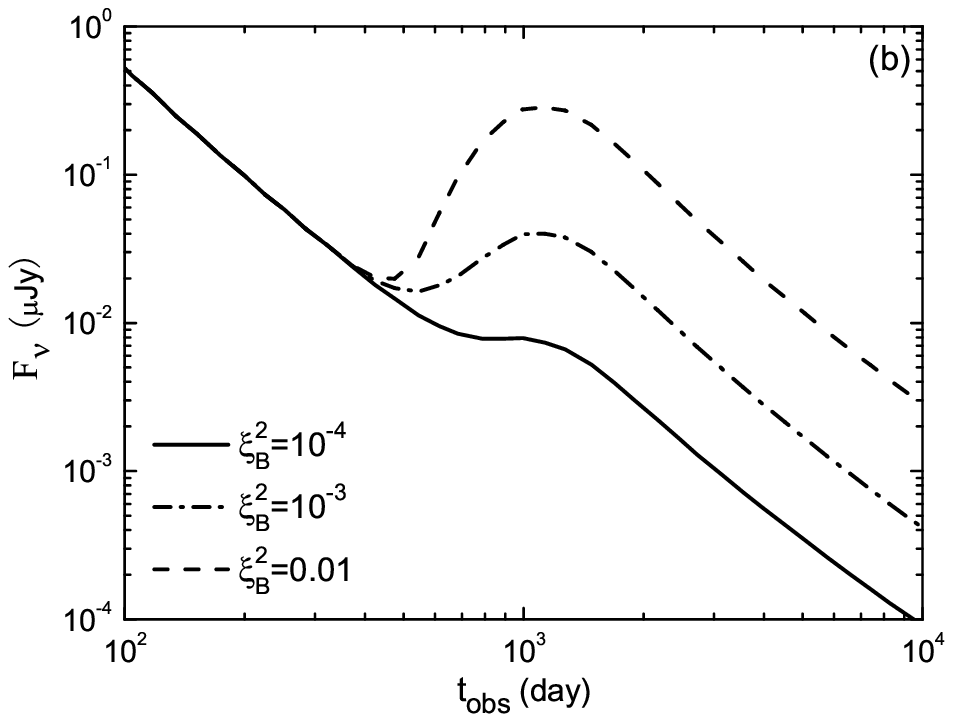}
   \includegraphics[width=0.35\paperwidth]{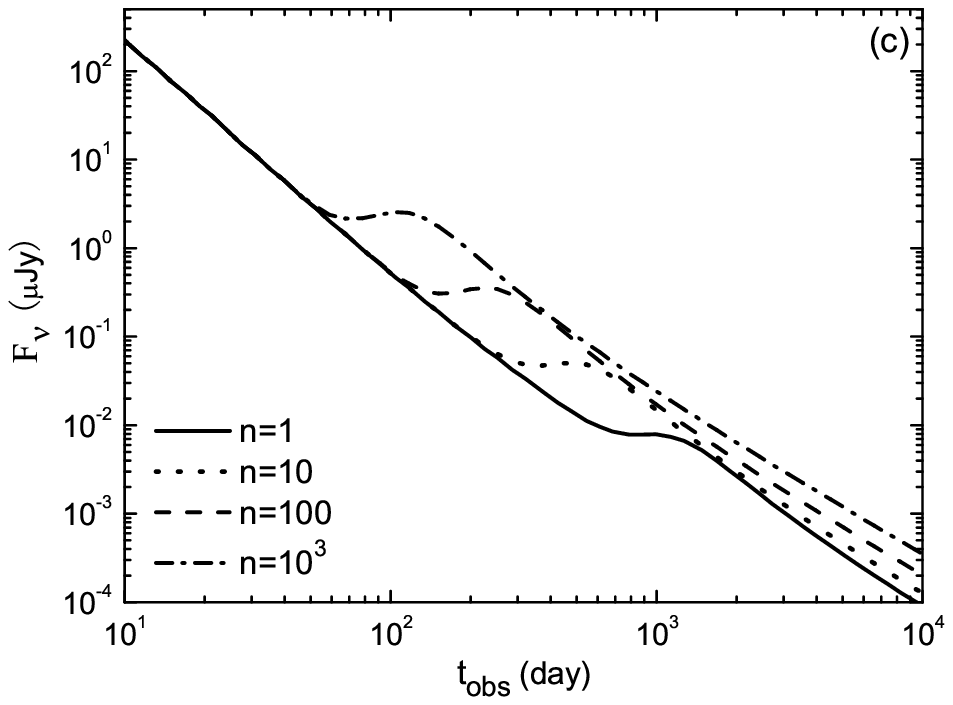} \includegraphics[width=0.35\paperwidth]{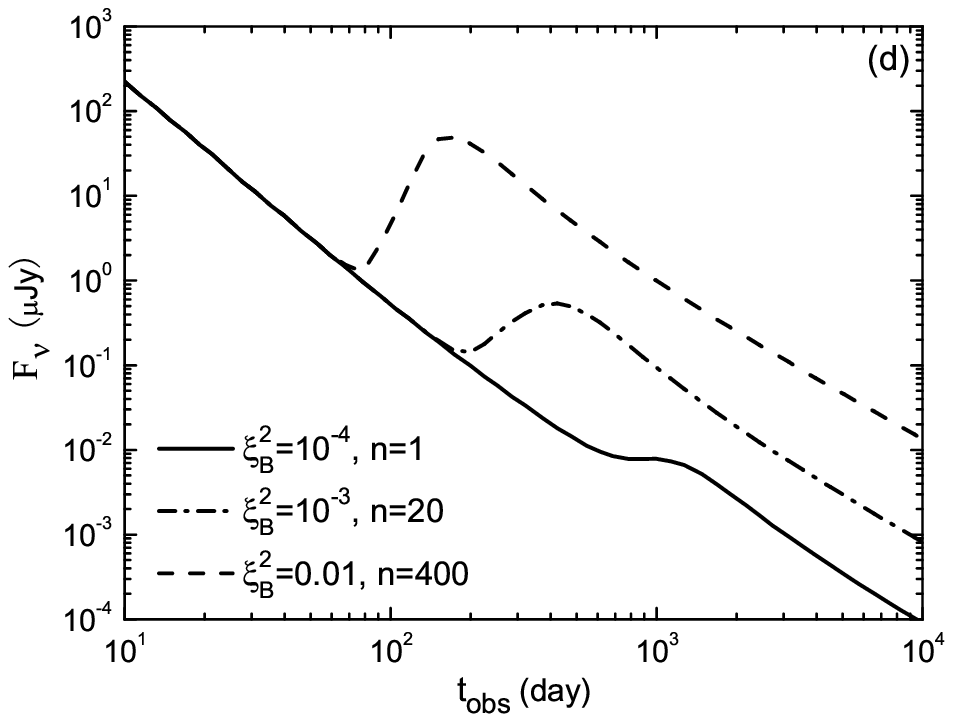}
   \par\end{center}
   \caption{8.46 GHz radio afterglow light curves of double-sided jets. In this figure, we
   assume that the parameters of the receding jet can be different from those of the forward jet.
   In each panel, the solid line is plotted under the ``standard'' condition, i.e., the parameters
   are completely the same for the twin jets (but note that we have evaluated $\xi_{\rm e}$ as 0.01
   and $\xi_{\rm B}^2$ as $10^{-4}$ here). For other light curves, one or two parameters
   are changed for the receding jet, to see its effect on the afterglows. }
   \label{fig7}
\end{figure*}

\label{lastpage}

\end{document}